\documentclass[twocolumn,showpacs,pra]{revtex4}

\usepackage[dvips]{graphicx,color}
\usepackage{delarray}
\usepackage{amsmath}
\usepackage{bm}

\newcommand{\braket}[2]{\langle {#1} | {#2} \rangle}
\newcommand{\ket}[1]{| { #1} \rangle}
\newcommand{\bra}[1]{ \langle {#1}  |}

\begin{document}
\title{Quantum repeaters and computation by a single module}
\author{Koji Azuma}
\email{azuma@qi.mp.es.osaka-u.ac.jp}
\affiliation{Department of Materials Engineering Science, Graduate School of Engineering Science,
Osaka University, Toyonaka, Osaka 560-8531, Japan}

\author{Hitoshi Takeda}
\affiliation{Department of Materials Engineering Science, Graduate School of Engineering Science,
Osaka University, Toyonaka, Osaka 560-8531, Japan}

\author{Masato Koashi}
\affiliation{Department of Materials Engineering Science, Graduate School of Engineering Science,
Osaka University, Toyonaka, Osaka 560-8531, Japan}

\author{Nobuyuki Imoto}
\affiliation{Department of Materials Engineering Science, Graduate School of Engineering Science,
Osaka University, Toyonaka, Osaka 560-8531, Japan}

\
\date{\today}

\begin{abstract}
We present a protocol of remote nondestructive parity measurement
(RNPM) on a pair of quantum memories.
The protocol works as a single module for key operations such as
entanglement generation, Bell measurement, parity check measurement,
and an elementary gate for extending one-dimensional cluster states.
The RNPM protocol is achieved by a simple combination of devices such
as lasers, optical fibers, beam splitters, and photon detectors.
Despite its simplicity, a quantum repeater composed of RNPM protocols
is shown to have
a communication time that scales sub-exponentially with the channel
length, and it can be
further equipped with entanglement distillation. With a reduction in
the internal losses,
the RNPM protocol can also be used for generating cluster states
toward measurement-based quantum communication. 
\end{abstract}
\maketitle

\begin{figure}[tb]
  \begin{center}
    \includegraphics[keepaspectratio=true,height=69mm]{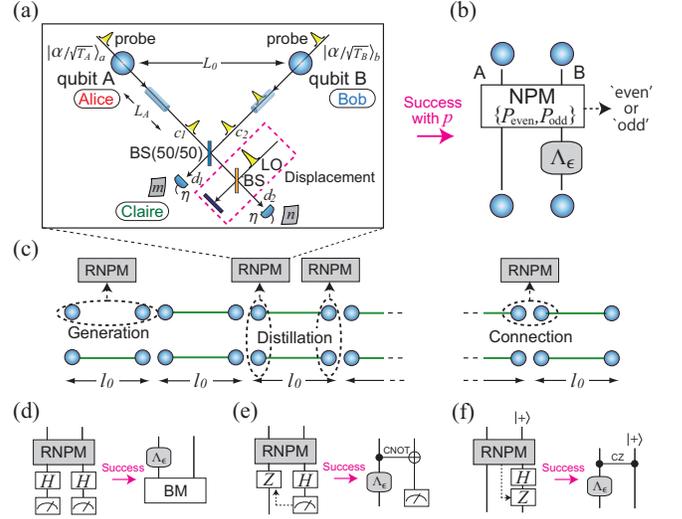}
  \end{center}
  \caption{(a) The RNPM protocol. (b) A circuit equivalent to the successful RNPM protocol, where a phase-flip channel $\Lambda_{\epsilon}(\hat{\rho}):=(1-\epsilon) \hat{\rho}+ \epsilon \hat{Z} \hat{\rho} \hat{Z}$ with phase error probability $\epsilon$ is applied as the penalty of photon losses. $\epsilon$ may depend on the outcome returned by photon detectors. In the lossless limit, the RNPM protocol works as the ideal NP measurement.
(c) Quantum repeaters based on the RNPM protocols. Applications of RNPM: (d) Bell measurement (BM), (e) parity check measurement, and (f) a gate for extending one-dimensional cluster state, where the measurement instrument means $\hat{Z}$-basis measurement and the dashed arrow implies the transmission of the measurement outcome.}
  \label{fig:fig1.eps}
\end{figure}

In quantum mechanics, measuring a property of a system inevitably
causes disturbance on its state. Hence an ideal measurement would be
the one that leaves the measured system with only as much disturbance
as is necessary. A simple nontrivial example of such a measurement is
the nondestructive parity (NP) measurement on two qubits $AB$, which
is the projection measurement to the subspace with even parity
spanned by $\{ \ket{00}_{AB},\ket{11}_{AB} \}$ and to the odd one spanned by $\{ \ket{01}_{AB},\ket{10}_{AB} \}$. When the
qubits are in state $\ket{\varphi}_{AB}$ initially, the unnormalized post-measurement state is
ideally either $\hat{P}_{\rm even}^{AB} \ket{\varphi}_{AB} $ or $\hat{P}_{\rm odd}^{AB} \ket{\varphi}_{AB}$, where $\hat{P}_{\rm even}^{AB}$ ($\hat{P}_{\rm odd}^{AB}$) is
the projection onto the even (odd) subspace. This measurement
provides a powerful tool when the two qubits are quantum memories
located far apart. For example, if we prepare each qubit in state
$\ket{+}:=(\ket{0}+\ket{1})/\sqrt{2} $, the NP measurement leaves the pair in maximally entangled state (Bell state)
$\sqrt{2} \hat{P}_{\rm even}^{AB} \ket{++}_{AB}= \ket{\Phi^{+}}_{AB} $ or $ \sqrt{2} \hat{P}_{\rm odd}^{AB} \ket{++}_{AB}=\ket{\Psi^{+}}_{AB} $, where $\ket{\Phi^{\pm}}_{AB} := (\ket{00}_{AB} \pm \ket{11}_{AB})/\sqrt{2}$ and $\ket{\Psi^\pm}_{AB} :=( \ket{01}_{AB} \pm \ket{10}_{AB})/\sqrt{2}$. Various other nontrivial
operations are also derived from the NP measurement (see Fig.~\ref{fig:fig1.eps} (d)-(f) below).

In this paper, we provide a simple protocol to implement the NP
measurement, which
we call remote nondestructive parity measurement (RNPM) protocol.
The protocol is based on an off-resonant coupling of light
pulses with the quantum memories, and it works even if the quantum
memories are distant. 
The deviation of the RNPM protocol from the
ideal NP measurement mainly comes from the loss in the optical channel, whose
transmission depends on its length $L$ as $\eta_L:=e^{-L/L_{\rm att}}$ with an attenuation length $L_{\rm att}$. 
This makes the RNPM protocol
probabilistic and noisy, but these imperfections behave in a
controlled way, even with the use of threshold detectors that cannot
distinguish one from two or more photons. As a result, the RNPM
protocol constitutes a viable module which can be singly used to
build a quantum repeater, in contrast to the other known repeater protocols \cite{B98, D01,B07,Z07,J07,S07,SSRG09,C06,L06,La06,L08,M08,A09,A10}.
Moreover, the local use of highly efficient RNPM protocols will also allow us to generate cluster states.

The requirement on the memory qubit for the RNPM protocol is
as follows. 
The qubit is assumed to allow us to apply phase flip
$\hat{Z}:=\ket{0}\bra{0} -\ket{1}\bra{1}$,
Hadamard gate $\hat{H}:=\ket{+}\bra{0} +\ket{-}\bra{1}$ with $\ket{-}:=\hat{Z} \ket{+}$, and $Z$-basis measurement. 
The qubit is also assumed to interact with an off-resonant laser pulse $a$ in a coherent state $\ket{\alpha}_a:=e^{-|\alpha|^2/2} \sum_{n=0}^\infty (\alpha^n/\sqrt{n!} ) \ket{n}_a$ according to a unitary operation $\hat{U}_\theta \ket{j} \ket{\alpha}_a = e^{ - i(-1)^j  \phi_\alpha /2}  \ket{j} \ket{\alpha e^{i (-1)^j \theta/2}}_a$ ($j=0,1$),  
where $\{\ket{n}_a\}$ are the number states of the mode $a$, $\phi_\alpha=\alpha^2 \sin \theta$, and $\theta$ is a fixed parameter for the strength of the interaction. 
Since this interaction is an off-resonant coupling based on a basic Hamiltonian -- Jaynes-Cummings Hamiltonian, 
it will be feasible with various qubits such as an individual $\Lambda$-type atom, a trapped ion, a single electron trapped in quantum dots, a nitrogen-vacancy (NV) center in a diamond with a nuclear spin degree of freedom, and a neutral donor
impurity in semiconductors \cite{L06,La06}.

We now describe our RNPM protocol in detail. Suppose that the qubits
$A$ and $B$ are respectively held by
Alice and Bob, who are distance $L_0$ apart [See
Fig.~\ref{fig:fig1.eps} (a)]. Claire is located in between, connected to Alice
and Bob with optical channels $a \to c_1$ and $b\to c_2$ with lengths
$L_A(\le L_0)$ and $L_B:=L_0-L_A$, respectively. Let $T_A:=\tau
\eta_{L_A}$ and $T_B:=\tau \eta_{L_B}$
be the overall transmittance of the channels, where $\tau$ stands for
the local loss.
The RNPM protocol proceeds as follows: 
(i) Alice (Bob) prepares pulse $a$ (pulse $b$) in a coherent state $\ket{\alpha/\sqrt{ T_{A}}}_a$ ($\ket{\alpha/\sqrt{ T_{B} }}_b$) with $\alpha\ge 0$, and let it interact with qubit $A$ (qubit $B$) by $\hat{U}_{\theta}$;
(ii) Alice (Bob) sends Claire the pulse $a$ (the pulse $b$) through the optical channel $a \to c_1$ ($b \to c_2$);
(iii) On receiving the pulses $c_1c_2$, Claire makes the pulses interfere by a half beam splitter;
(iv) On the mode receiving the constructive interference, Claire applies displacement operation $\hat{D}( - \sqrt{2} \alpha \cos (\theta/2))$ by using a local oscillator (LO);
(v) Claire counts photons of the output modes $d_1d_2$ by two photon detectors, and she announces the outcome $(m,n)$; 
(vi) If $m+n$ is odd, Bob applies phase flip $\hat{Z}$ to qubit $B$.
Events with $m>0$ and $n=0$ ($m=0$ and $n>0$) indicates outcome `odd' (`even'), which are regarded as the success events of this protocol.

To see the back actions in the success events,
we use the fact that the RNPM protocol works equivalently 
if we omit step (iv) and replace step (i) with the following:
(i') After making pulse $a$ (pulse $b$) in a coherent state $\ket{\alpha/\sqrt{ T_{A} } }_a$ ($\ket{\alpha/\sqrt{ T_B }}_b$) interact with qubit $A$ (qubit $B$), Alice (Bob) applies displacement operation $\hat{D}(-  (\alpha /\sqrt{ T_{A}}) \cos (\theta/2))$ ($\hat{D}(-  (\alpha /\sqrt{ T_{B}}) \cos (\theta/2))$) on the pulse.
In this protocol, through steps (i')-(iii), qubits $AB$ are transformed as
\begin{equation}
\begin{split}
\ket{00}_{AB} \stackrel{{\rm (i')}}{\to}&    \ket{00}_{AB} \ket{i \beta_A}_a \ket{i \beta_B}_{b}  {\to} \ket{00}_{AB} \ket{0}_{d_1} \ket{i \sqrt{2} \beta}_{d_2} , \\
\ket{01}_{AB} \stackrel{{\rm (i')}}{\to}&    \ket{01}_{AB} \ket{i\beta_A}_a \ket{-i \beta_B}_{b}  {\to} \ket{01}_{AB} \ket{-i \sqrt{2}  \beta}_{d_1}  \ket{0}_{d_2} , \\
\ket{10}_{AB} \stackrel{{\rm (i')}}{\to}&    \ket{10}_{AB} \ket{-i  \beta_A}_a \ket{i \beta_B}_{b}  {\to} \ket{10}_{AB} \ket{i \sqrt{2} \beta}_{d_1} \ket{0}_{d_2} , \\
\ket{11}_{AB} \stackrel{{\rm (i')}}{\to}&    \ket{11}_{AB} \ket{-i \beta_A}_a \ket{-i \beta_B}_{b}  {\to} \ket{11}_{AB} \ket{0}_{d_1} \ket{-i  \sqrt{2} \beta}_{d_2} , \label{eq:id'}
\end{split}
\end{equation}
where $\beta:=\alpha \sin (\theta/2) $ and $\beta_X := \beta/ \sqrt{T_X}$ $(X=A,B)$. 
Since this protocol does not use LO after (i'), we are allowed to assume that
the total number $k$ of photons in modes $ab$ was measured after step (i'),
without affecting the protocol at all.

We start with the ideal case where $T_{A}=T_{B}=1$ and the detectors at modes $d_1d_2$ are the ideal photon-number-resolving detectors.
Then, the $k$ photons in modes $ab$ are preserved throughout steps (ii) and (iii), which leads to $m+n=k$. Combined with Eq.~(\ref{eq:id'}), this suggests that all the $k$ photons are captured by one of the detectors.
Hence, if photon detector $d_1$ ($d_2$) announces the arrival of $k(>0)$ photons, from $\braket{k}{0}=0$ and $\braket{k}{-i\sqrt{2}\beta}=(-1)^k \braket{k}{i\sqrt{2}\beta}$, we see that the back action of the RNPM protocol is $\hat{P}^{AB}_{\rm odd}$ ($\hat{P}^{AB}_{\rm even}$) after Bob's phase flip at step (vi).

We can easily describe the back actions of the RNPM protocol with practical channels and detectors,
as long as the dark counting are negligible, namely, $\ket{0}_{d_1}$
always produces $m=0$.
This guarantees that the success outcome still gives the correct parity, but
$l:=m+n$ is no longer equal to $k$. Since the back action depends only on $(-1)^k$,
we see the following. If $l \equiv  k\; ({\rm mod }\; 2)$,
the final state is the same as the ideal case. Otherwise, the final
state suffers from a phase flip error $\hat{Z}^B$.
This observation means that
the success probability $p$ and
the phase error probability $\epsilon$ (conditioned on the success)
are solely determined from
the joint probability $Q(k,l)$ as follows,
\begin{equation}
p= \sum_{l \ge 1} \chi_l^+,\;\; \epsilon= \frac{1}{2p} \sum_{l\ge 1} ( \chi^+_l -\chi^-_l ), \label{eq:p-e}
\end{equation}
with $\chi_{l}^\pm := \sum_{k} (\pm 1)^{k-l} Q(k,l)$.

Let us derive the explicit forms of $(p,\epsilon)$ with various types of
detectors with quantum efficiency $\eta$.
Here we consider the case $T_A=T_B (=T)$ for simplicity, and
the general cases are treated in Appendix A. Since $k$ is the total
number of photons in two coherent states with amplitude $i\beta/\sqrt{T}$,
it follows the Poissonian distribution
$P_{\lambda}(k):=(e^{-\lambda} \lambda^k)/k!$ with $\lambda=2\beta^2/T$.
When photon-number resolving detectors are used, $l=m+n$ is the number of
photons that has passed through a channel with transmittance $\eta T$.
Hence we have $Q(k,l)=Q_{\infty}(k,l) := B_{\eta T }(l|k) P_{2 \beta^2/T}(k)$
with a binomial distribution $B_{p}(l|k):= [p^{l} (1-p)^{k-l}  k!]/[l! (k-l)!] $.
Using Eq.~(\ref{eq:p-e}), we have $p(\beta)=1-e^{-2 \beta^2 \eta}$ and
$\epsilon(\beta, T)= ( 1-e^{- 2 \beta^2 \eta [ 2(\eta T)^{-1}-2 ] } ) /2$.
When we use single photon detectors, we are informed of detection
of exactly one photon. Hence we have
$Q(k,1)=Q_{\infty}(k,1)$ and $Q(k,0)=Q_{\infty}(k,0)+\sum_{l \ge 2}
Q_{\infty}(k,l)$,
leading to $p(\beta)=P_{2 \eta \beta^2}(1)$ and
$\epsilon(\beta, T)=( 1-e^{- 2 \beta^2 \eta [ 2( \eta T)^{-1}-2 ] } ) /2$.
When threshold detectors are used, from
$Q(k,1)=\sum_{l\ge 1} Q_{\infty}(k,l)$, we obtain
$p(\beta)=1-e^{-2 \beta^2 \eta}$ and
$\epsilon(\beta, T)=( 1-e^{- 2 \beta^2 \eta [ 2  (\eta T)^{-1}- 1  ] } ) /2$.

As seen in the above examples, the success probability $p$ and
the phase error probability $\epsilon$ of the RNPM protocol are
under a trade-off relation, which is controllable by $\beta$, namely
by $\alpha$.
For a fixed $L_0$, the choice of $L_A=L_B=L_0/2$ gives the best
performance of $(p,\epsilon)$. On the other hand, the choice $L_A=L_0$
has a technical merit in stabilizing the relative phase between
pulses $c_1$ and $c_2$.
The RNPM protocol can be also used for interacting quantum memories located in
a single site, in which case $L_0$ is nearly zero and the local loss
$\tau$ determines
the trade-off relation. We describe various applications of the RNPM
protocol below.

Long-distance quantum communication over lossy channels:
The goal here is to share an entangled pair of qubits between two end
stations separated by distance $L$.
With direct transmission of single photons, the communication time
would increase exponentially with distance $L$ according to $e^{L/L_{\rm att}}$.
Disposition of relaying stations with quantum memories helps to avoid
the exponential increase by using
a quantum repeater protocol \cite{B98}. Let us see how a repeater
protocol is built up from the
RNPM protocol. Suppose that the stations are placed at $l_0:=L/2^n$
intervals (see Fig.~\ref{fig:fig1.eps} (c)).
Each station has at least two qubits.

The first step is entanglement generation between neighboring
stations separated by $l_0$.
The RNPM protocol is applied to the two qubits in state
$\ket{+}\ket{+}$, and is repeated
until it is successful. Assuming the time $l_0/c$ for each trial,
it takes time $(l_0/c) p(\beta_g)^{-1}$ on average, and the Bell
state is produced
with phase error probability $\epsilon_0:=\epsilon(\beta_g, \tau \eta_{l_0/2})$.
Here we consider the case with $L_A=L_B=l_0/2$ for simplicity of the notations.
(The cases with $L_A=l_0$ are found in Appendix B)

\begin{figure}[tb]
  \begin{center}
    \includegraphics[keepaspectratio=true,height=30mm]{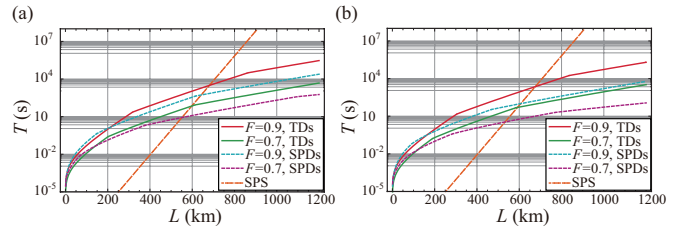}
  \end{center}
  \caption{The minimum time $T$ needed to generate entanglement with $F=0.9$, $0.7$ over distance $L$ under the use of threshold detectors (TDs) and single photon detectors (SPDs): (a) $\tau=0.95$ and $\eta=0.9$; (b) $\tau=0.98$ and $\eta=0.95$. $c=2 \times 10^8$ m/s, $L_{\rm att}=22$ km. The direct transmission time $(f \eta T_L)^{-1} $ of the photon from $10$ GHz ($f=10^{10}$) single photon source (SPS) is also shown as a reference.}
  \label{fig:th-m-time.eps}
\end{figure}

Next, the repeater protocol proceeds to entanglement connection \cite{Z93}.
Suppose that two stations separated by $2^{j} l_0$ ($j=0,1,\ldots,
n-1$) can share a qubit pair in the Bell state
with phase error probability $\epsilon_j$ and with average time
$t_j$. After creating two such pairs connecting
three stations, the middle one executes the Bell measurement by locally applying
the RNPM protocol as in Fig.~\ref{fig:fig1.eps} (d), which succeeds
with probability $p(\beta_s)$ and produces
entangled qubits $2^{j+1} l_0$ apart.
Adding up the contribution
of the phase errors in the two initial pairs and in the Bell
measurement, we have
$1-2 \epsilon_{j+1} = (1-2 \epsilon_{j})^2 (1-2 \epsilon(\beta_s, \tau ))$.
Since it approximately takes time $(3/2) t_j$ per trial \cite{SSRG09}, we have
$t_{j+1} \sim (3/2) t_j p(\beta_s)^{-1}$ for the average time for success.
Solving these recursive relations, we see that the average total time
$T=t_n$ is approximately written as
\begin{equation}
T \sim \frac{l_0}{c} \left( \frac{3}{2} \right)^{ \log_2 (L/l_0)}
p(\beta_g)^{-1}  p(\beta_s)^{-\log_2 (L/l_0)},
\end{equation}
and the final state is $\hat{\rho}^{AB}= F
\ket{\Phi^+}\bra{\Phi^+}_{AB} + (1-F) \ket{\Phi^-}\bra{\Phi^-}_{AB} $ with
\begin{equation}
2F-1=(1-2 \epsilon (\beta_g, \tau \eta_{l_0/2}))^{L/l_0} (1-2
\epsilon (\beta_s, \tau))^{L/l_0 -1}.
\end{equation}
For large $L$, it should be chosen as $\beta_g^2 \sim \beta_s^2 \sim
{\cal O} (l_0/L)$.
Then, noticing that $p(\beta) \sim {\cal O}( \beta^2)$ and $\epsilon
(\beta,T) \sim {\cal O} (\beta^2)$ hold regardless of the types of
the photon detectors, we have $F \sim {\cal O}(1)$ and $T \sim {\cal
O} ( ( 3/ 2)^{ \log_2 (L/l_0) } ( L/ l_0)^{ \log_2 (L/l_0)+1 })$.
Hence, $T$ increases only sub-exponentially with $L$.
We also numerically optimized $T$ for fixed values of final fidelity
$F$ and the distance $L$, which are shown in Fig.~\ref{fig:th-m-time.eps}.

We stress that the generated state $\hat{\rho}^{AB}$ includes only one-type of error, which is a good property for quantum communication.
For example, for the state $\hat{\rho}^{AB}$, the formula of secure key rate of the entanglement-based protocol \cite{E91,BBM92} is proportional to $1-h(F)$ with the binary entropy function $h(x):=-x \log_2 x-(1-x) \log_2 (1-x)$, which implies that the secret key is distillable for any $F>1/2$.

Entanglement distillation: While the optical losses considered above
are the dominant obstacle in long-distance
communication, other types of small noises will be also present.
Entanglement distillation not only helps to counter such general
errors, but also reduces the scaling of
the communication time to be polynomial in distance $L$ \cite{B98}.
In a simple method of distillation called the recurrence method \cite{B96},
Alice and Bob first transform each pair of qubits locally into the
so-called Werner state while keeping
the fidelity $F$ to a Bell state. Suppose that they have two such
pairs $A_1B_1$ and $A_2 B_2$ with $F>1/2$.
Alice applies C-NOT gate on her qubit $A_1$ as the control and on
$A_2$ as the target, and measures $A_2$ on
${Z}$-basis (the whole process is called parity check measurement).
Bob also applies the same measurement on his qubits.
Their outcomes will agree with a probability $P_{\rm rec}(F)$,
and then the remaining pair $A_1B_1$ will have an improved fidelity.

\begin{figure}[tb]
  \begin{center}
    \includegraphics[keepaspectratio=true,height=29mm]{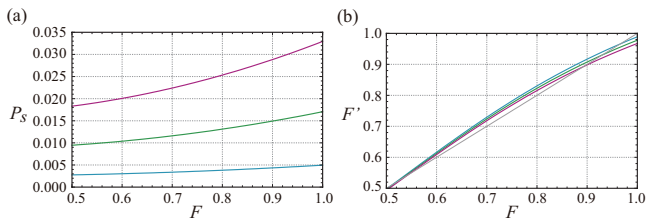}
  \end{center}
  \caption{For $\beta^2=0.04,0.08,0.12$, the efficiencies of the recurrence method based on the RNPM protocols with single photon detectors as a function of fidelity $F$ of the Werner states to a Bell state; (a) The success probability $P_s=P_s^d (P_{2 \eta \beta^2 } (1))^2$, (b) The fidelity $F'$ of the left qubit. $\tau=0.98$ and $\eta=0.95$.}
  \label{fig:Distill-Werner-th-P-F.eps}
\end{figure}

Since the outcome of each party is the parity of the two qubits, it
can also be obtained via the RNPM protocol.
In addition, if the RNPM protocol succeeds, by subsequently measuring
$A_2$ on $X$ basis to produce outcome
$x$ and then by applying $\hat{Z}^x$ on $A_1$, the post-measurement
state of $A_1$ is also simulated
except the phase error $\epsilon(\beta,\tau)$ [see
Fig.~\ref{fig:fig1.eps} (e)]. The overall success probability is
$P_s:=P_{\rm rec}(F) p^2(\beta)$, which is in a trade-off relation
with the fidelity $F'$ of the final state and
is controllable through $\beta$. In Fig.~\ref{fig:Distill-Werner-th-P-F.eps},
we give numerical examples with single photon detectors.

Generation of cluster states:
One promising way for implementing quantum computing is the so-called
measurement-based quantum computation, where computation proceeds
with sequential one-qubit measurements on a system in a highly
entangled state -- the cluster state \cite{RB00,BBDR09}. The
addressing of individual qubits is easier when they are located not
so close to each other. Such a sparse configuration also helps to
reduce correlated errors from the environment. In this case, the RNPM protocol
works as an entangler for qubits that are not in close proximity. In
fact, the gate shown in Fig.~\ref{fig:fig1.eps} (f) can be
used for extending one-dimensional cluster states, and the parity
check measurement in Fig.~\ref{fig:fig1.eps} (e)
can be used for fusing two cluster states \cite{BR05,DR05}.
The combination of these two types of gates enables us to
build up a large cluster state.
Hence, with future development of good detectors and
reduction of internal losses, the RNPM protocol will also work as a
tool for implementing quantum computing.

We have proposed a versatile protocol, called the RNPM protocol, for
measuring the parity of two separated qubits
in a non-destructive way. The performance of the RNPM protocol is
simply related to the optical loss and the
characteristics of photon detectors.
We have shown that, even with threshold detectors, the protocol can
be used as a module to build up quantum repeaters
for long-distance quantum communication. Efficient single photon
detectors will allow us to equip the repeaters with
entanglement distillation, a countermeasure against arbitrary types
of noises. With further improvement of the performance,
more general quantum computation will be brought within the scope
through the generation of cluster states via
the RNPM protocol. We believe that the existence of such a
versatile protocol puts a renewed interest in developing
efficient photon detectors and quantum memories off-resonantly coupled to light.

We would like to thank Naoya Sota for valuable discussions.
We acknowledge the support
of a MEXT Grant-in-Aid for Scientific Research on
Innovative Areas 21102008, a MEXT Grant-in-Aid for
the Global COE Program, JSPS Grant-in-Aid for Scientific
Research (C) 20540389. K.A. is
supported by JSPS.

\appendix


\appendix
\onecolumngrid
\section{The performance of the RNPM protocol}
Here, for arbitrary values of $T_A$ and $T_B$, we derive the
performance $(p,\epsilon)$ of the RNPM protocol
with various types of detectors.
As shown in the main body of this paper, the performance is
determined by calculating the joint probability $Q(k,l)$
with which modes $ab$ have $k$ photons in total and the arrival of
$l$ photons is announced by photon detectors
$d_1 d_2$ in total.
Let $k_a$ and $k_b$ be the numbers of photons in modes $a$ and $b$, respectively.
Since mode $a$ is in a coherent state with amplitude $i \beta
/\sqrt{T_A}$, $k_a$ follows the Poissonian distribution
$P_{\beta^2/T_A}(k_a)$ with
$P_{\lambda}(k):=(e^{-\lambda} \lambda^k)/k!$.
Similarly, $k_b$ obeys the Poissonian distribution $P_{\beta^2/T_A}(k_b)$.

Suppose that we use photon-number-resolving detectors with quantum
efficiency $\eta$ for the detectors $d_1$ and $d_2$.
Each of the $k_a$ photons will then be detected with probability
$\eta T_A$. Hence, the probability of detecting $l_a$ photons among
$k_a$ photons in mode $a$
is given by $B_{\eta T_A} (l_a |k_a) P_{\beta^2 /T_A} (k_a)$, where
$B_{p}(l|k):=[p^l (1-p)^{k -l} k!]/[l! (k-l)!]$ is the binomial distribution.
Similarly, the probability of detecting $l_b$ photons among $k_b$
photons in mode $b$
is given by $B_{\eta T_B} (l_b |k_b) P_{\beta^2 /T_B} (k_b)$.
Since $Q(k,l)$ is given by the sum of all probabilities with
$k=k_a+k_b$ and $l=l_a+l_b$, we have
\begin{align}
Q(k,l)=& Q_{\infty}(k,l) := \sum_{l_a=0}^l \sum_{k_a=l_a}^{l_a+(k-l)}
B_{\eta T_A} (l_a |k_a) P_{\beta^2 /T_A} (k_a) B_{\eta T_B} (l-l_a
|k-k_a) P_{\beta^2 /T_B} (k-k_a) \nonumber \\
=& e^{-\beta^2 (\frac{1}{T_A}+\frac{1}{T_B})}  (\eta \beta^2 )^l \sum_{l_a=0}^l
  \frac{1}{l_a!(l-l_a)!} \sum_{k_a=l_a}^{l_a+(k-l)}   \frac{\left(
\frac{1-\eta T_A}{T_A} \beta^2  \right)^{k_a-l_a}  \left(
\frac{1-\eta T_B}{T_B} \beta^2  \right)^{k-l-(k_a-l_a)} }{(k_a-l_a)!
[k-l-(k_a-l_a)]! } \nonumber \\
=& e^{-\beta^2 (\frac{1}{T_A}+\frac{1}{T_B})}  (\eta \beta^2 )^l
 \frac{1}{(k-l)!} \left[ \left( \frac{1-\eta T_A}{ T_A} +
\frac{1-\eta T_B}{T_B}  \right) \beta^2   \right]^{k-l}
\sum_{l_a=0}^l  \frac{1}{l_a!(l-l_a)!}  \nonumber  \\
=& \frac{e^{-(\frac{1}{T_A} +\frac{1}{T_B})\beta^2}}{l! (k-l)!} (2
\beta^2 \eta)^l \left[ \left( \frac{1-\eta T_A}{T_A}+\frac{1-\eta
T_B}{T_B} \right) \beta^2 \right]^{k-l},
\end{align}
where we used the binomial theorem
\begin{equation}
(a+b)^n= \sum_{m=0}^n \frac{n!}{m! (n-m)!} a^m b^{n-m}
\end{equation}
for any $a,b \in {\bm R}$ and $n \in {\bm N}$.
From the expression of $Q(k,l)$, $\chi^{\pm}_l$ are calculated to be
\begin{align}
&\chi_l^{+} = \sum_{k} Q(k,l)= \sum_{k=l}^\infty Q_{\infty}(k,l) =
\frac{(2 \beta^2 \eta )^l}{l!} e^{ -2 \beta^2 \eta }, \label{eq:chi-1} \\
&\chi_l^{-} = \sum_{k} (-1)^{k-l} Q(k,l) = \sum_{k=l}^\infty
(-1)^{k-l} Q_{\infty}(k,l)= \frac{(2 \beta^2 \eta )^l}{l!} e^{- 2
\beta^2 \eta  \left( \frac{1}{\eta T_A}+\frac{1}{\eta T_B} -1
\right) }, \label{eq:chi-2}
\end{align}
by noting $e^x =\sum_{m=0}^\infty x^m/m!$.
Hence, the success probability $p$ and the phase error probability
$\epsilon$ of the RNPM protocol with photon-number-resolving detectors are
\begin{align}
&p(\beta)=\sum_{l \ge 1} \chi^+_l = \sum_{l=1}^\infty \chi^+_l =
1-e^{-2  \beta^2 \eta}, \label{eq:p-n}  \\
&\epsilon(\beta, T_A,T_B)= \frac{1}{2 p} \sum_{l\ge 1} (\chi_l^+
-\chi_l^-) = \frac{1}{2 p} \sum_{l=1}^\infty (\chi_l^+ -\chi_l^-)
=\frac{1}{2} \left( 1- e^{-2  \beta^2  \eta \left( \frac{1}{\eta T_A}
+\frac{1}{\eta T_B} - 2 \right) }  \right). \label{eq:e-n}
\end{align}
Note that the above expressions are reduced to the ones in the main
body of the paper for $T_A=T_B(=T)$.
By substituting $T_A=\tau \eta_{L_A}= \tau e^{-L_A/L_{\rm att}}$ and
$T_B=\tau \eta_{L_0-L_A}=\tau e^{-{(L_0-L_A)/L_{\rm att}}}$ into
Eqs.~(\ref{eq:p-n}) and (\ref{eq:e-n}), one can easily confirm that,
for a fixed $L_0$, the choice of $L_A=L_B=L_0/2$ gives the best performance.
In other words, the RNPM protocol works best when Claire is located
at the middle point between Alice and Bob.

\subsection{Use of single photon detectors}
Here we assume the use of single photon detectors with quantum
efficiency $\eta$, which announce the detection of photons only when
receiving exactly one photon.
In this case, $Q(k,l)$ is described by
\begin{align}
& Q(k,1)=Q_{\infty} (k,1), \\
& Q(k,0)=Q_{\infty}(k,0) + \sum_{l\ge 2} Q_{\infty} (k,l) .
\end{align}
Then, $\chi^{\pm}_1$ are calculated to be
\begin{align}
&\chi_1^{+} = \sum_{k} Q(k,1)= \sum_{k=1}^\infty Q_{\infty}(k,1) =2
\beta^2 \eta e^{- 2 \beta^2 \eta}, \\
&\chi_1^{-} = \sum_{k} (-1)^{k-1} Q(k,1) = \sum_{k=l}^\infty
(-1)^{k-1} Q_{\infty}(k,1)= 2 \beta^2 \eta  e^{- 2 \beta^2 \eta
\left( \frac{1}{\eta T_A}+\frac{1}{ \eta T_B} -1 \right)  }
\end{align}
from the last equations in Eqs.~(\ref{eq:chi-1}) and (\ref{eq:chi-2}).
Hence, we conclude
\begin{align}
&p(\beta)=\sum_{l \ge 1} \chi^+_l = \chi_1^{+} = 2 \beta^2 \eta e^{-
2 \beta^2 \eta}, \label{eq:sing-p}  \\
&\epsilon(\beta,T_A,T_B)= \frac{1}{2 p} \sum_{l\ge 1} (\chi_l^+
-\chi_l^-) =\frac{1}{2 p} (\chi_1^+ -\chi_1^-)  = \frac{1}{2} \left(
1- e^{-2  \beta^2  \eta \left( \frac{1}{\eta T_A} +\frac{1}{\eta T_B}
- 2 \right) }  \right). \label{eq:sing-e}
\end{align}

\subsection{Use of threshold detectors}
Here we consider the case of threshold detectors with quantum efficiency $\eta$.
Since this type of detectors click only when receiving nonzero photons,
we have
\begin{align}
&Q(k,1)=\sum_{l \ge 1} Q_{\infty}(k,l),   \\
&Q(k,0)= Q_{\infty} (k,0),
\end{align}
From this, $\chi^{\pm}_1$ are calculated to be
\begin{align}
\chi_1^{+} =& \sum_{k} Q(k,1)=  \sum_{k=1}^\infty  \sum_{l=1}^{k}
Q_{\infty}(k,l)= \sum_{l=1}^\infty \sum_{k=l}^{\infty}
Q_{\infty}(k,l) = \sum_{l=1}^{\infty}   \frac{(2\beta^2 \eta)^l }{l!}
e^{-2 \beta^2 \eta}= 1-e^{-2 \beta^2 \eta}, \\
\chi_1^{-} =& \sum_{k} (-1)^{k-1} Q(k,1) = \sum_{k=1}^\infty
\sum_{l=1}^k (-1)^{k-1}   Q_{\infty}(k,l)=  \sum_{l=1}^\infty
(-1)^{l-1} \sum_{k=l}^{\infty} (-1)^{k-l}  Q_{\infty}(k,l) \nonumber \\
=& \sum_{l=1}^\infty  (-1)^{l-1} \frac{(2 \beta^2 \eta )^l}{l!}
e^{-\left[ \left( \frac{2-\eta T_A}{T_A}+\frac{2-\eta T_B}{T_B}
\right) \beta^2 \right]} = (1-e^{-2 \beta^2 \eta}) e^{- 2 \beta^2
\eta  \left( \frac{1}{\eta T_A}+\frac{1}{\eta T_B} -1 \right) }
\end{align}
from the last equations in Eqs.~(\ref{eq:chi-1}) and (\ref{eq:chi-2}).
Hence, the success probability $p$ and the phase error probability $\epsilon$ are
\begin{align}
&p(\beta)=\sum_{l \ge 1} \chi^+_l =  \chi^+_1 = 1-e^{-2  \beta^2
\eta}, \label{eq:th-p} \\
&\epsilon(\beta, T_A,T_B)= \frac{1}{2 p} \sum_{l\ge 1} (\chi_l^+
-\chi_l^-) = \frac{1}{2 p} (\chi_1^+ -\chi_1^-) =\frac{1}{2} \left(
1-  e^{- 2 \beta^2 \eta  \left( \frac{1}{\eta T_A}+\frac{1}{\eta T_B}
-1 \right) } \right). \label{eq:th-e}
\end{align}

\section{The performance of long-distance quantum communication over
lossy channels}

\begin{figure}[tb]
  \begin{center}
    \includegraphics[keepaspectratio=true,height=33mm]{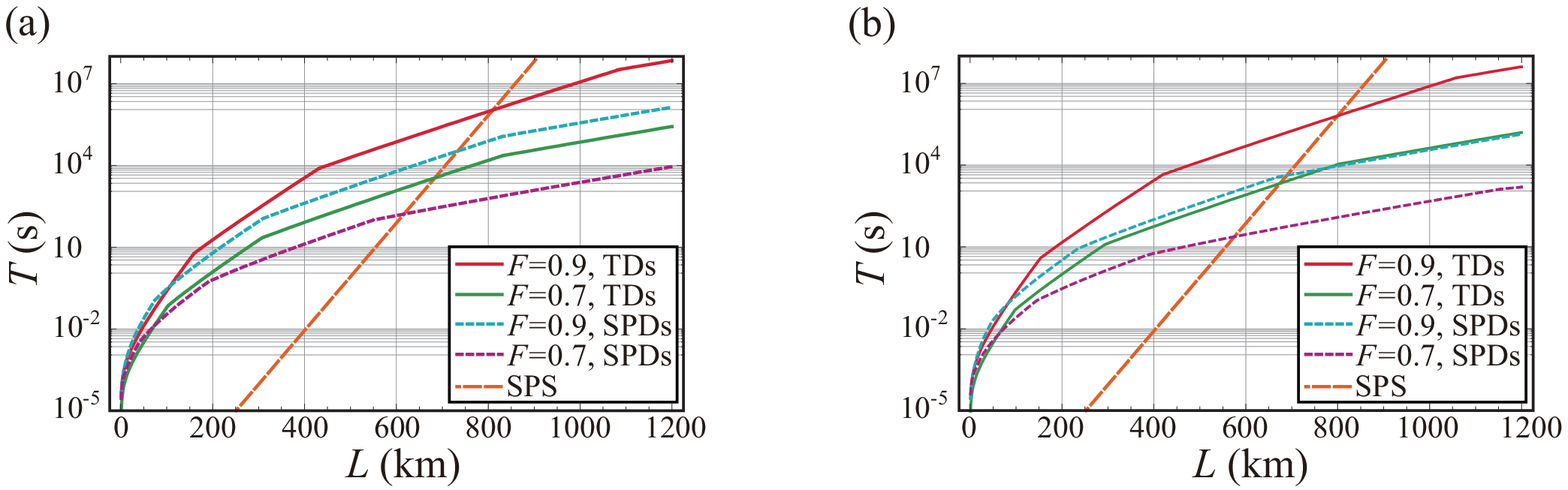}
  \end{center}
  \caption{For $L_A=l_0$ and $L_B=0$, the minimum time $T$ needed to generate entanglement with $F=0.9$, $0.7$ over distance $L$ under the use of threshold detectors (TDs) and single photon detectors (SPDs): (a) $\tau=0.95$ and $\eta=0.9$; (b) $\tau=0.98$ and $\eta=0.95$. $c=2 \times 10^8$ m/s, $L_{\rm att}=22$ km. The direct transmission time $(f \eta T_L)^{-1} $ of the photon from $10$ GHz ($f=10^{10}$) single photon source (SPS) is also described as a reference.}
  \label{fig:time-for-appendix.eps}
\end{figure}

Although the RNPM protocol has the best performance when Claire is
halfway between Alice and Bob,
the choice with $L_B=0$ (Claire's task is executed by Bob) is also
worth mentioning since
the stabilization of the relative phase between
pulses $c_1$ and $c_2$ is easier.
Here we calculate the performance of quantum repeaters with this technical merit.
More precisely, we assume the use of the RNPM protocols with
$L_A=l_0=L/2^n$ and $L_B=0$ for the entanglement generation. 
In this case, the average total time $T$ and the fidelity $F$ are described by 
\begin{align}
T \sim & \frac{l_0}{c} \left( \frac{3}{2} \right)^{\log_2 (L/l_0)} p(\beta_g)^{-1} p(\beta_s)^{-\log_2 (L/l_0)} \label{T}, \\ 
F= & \frac{1+ (1-2 \epsilon (\beta_g, \tau \eta_{l_0},\tau ) )^{L/l_0} (1-2 \epsilon (\beta_s, \tau,\tau) )^{L/l_0-1}}{2} .\label{F}
\end{align}
By substituting Eqs.~(\ref{eq:sing-p}) and (\ref{eq:sing-e}) [or Eqs.~(\ref{eq:th-p}) and (\ref{eq:th-e})] into these equations, 
we numerically optimized $T$ for fixed values of final fidelity $F$ and the distance $L$, which are shown in Fig.~\ref{fig:time-for-appendix.eps}.

\twocolumngrid

\end{document}